\documentstyle[12pt]{article}

\begin{document}


\font\sansf=cmss12


\newcommand{\inner}[2]{\langle #1\mid #2\rangle}
\newcommand{\bra}[1]{\langle #1 |}
\newcommand{\ket}[1]{| #1\rangle}


\newcommand{\GG}{{\cal G}}
\newcommand{\JJ}{{\cal J}}
\newcommand{\Ft}{{\cal F}}
\newcommand{\Ct}{{\cal C}}
\newcommand{\St}{{\cal S}}
\newcommand{\Tt}{{\cal T}}
\newcommand{\Kt}{{\cal K}}
\newcommand{\Vt}{{\cal V}}
\newcommand{\Wt}{{\cal W}}

\newcommand{\DD}{{\cal D}}
\newcommand{\HH}{{\cal H}}
\newcommand{\KK}{{\cal K}}
\newcommand{\LL}{{\cal L}}
\newcommand{\MM}{{\cal M}}
\newcommand{\NNN}{{\cal N}}
\newcommand{\PP}{{\cal P}}
\newcommand{\Pt}{{\cal P}}
\newcommand{\Qt}{{\cal Q}}
\newcommand{\SSS}{{\cal S}}

\newcommand{\square}{\vrule height 1.5ex width 1.2ex depth -.1ex }


\newcommand{\II}{\leavevmode\hbox{\small1\kern-3.8pt\normalsize1}}

\newcommand{\inbar}{\,\vrule height1.5ex width.4pt depth0pt}
\newcommand{\CC}{\relax{\hbox{$\inbar\kern-.3em{\rm C}$}}}
\newcommand{\RR}{{\rm I\! R}}
\newcommand{\NN}{{\rm I\! N}}
\newcommand{\ZZ}{\hbox{\sansf Z\kern-0.4em Z}}


\newcommand{\Coinfd}{C_0^\infty(\RR^d\backslash\{ 0\})}
\newcommand{\Coinf}[1]{C_0^\infty(\RR^{#1}\backslash\{ 0\})}
\newcommand{\Coin}{C_0^\infty(0,\infty)}


\newtheorem{Thm}{Theorem}[section]
\newtheorem{Def}[Thm]{Definition}
\newtheorem{Lem}[Thm]{Lemma}
\newtheorem{Prop}[Thm]{Proposition}
\newtheorem{Cor}[Thm]{Corollary}


\renewcommand{\theequation}{\thesection.\arabic{equation}}
\newcommand{\sect}[1]{\section{#1}\setcounter{equation}{0}}


\newcommand{\hg}{h_{\rm gpi}}

\newcommand{\diag}{{\rm diag}\,}
\newcommand{\sgn}{{\rm sgn}\,}
\newcommand{\sig}{{\rm sig}\,}
\newcommand{\Ran}{{\rm Ran}\,}
\newcommand{\rank}{{\rm rank}\,}
\newcommand{\Span}{{\rm span}\,}

\begin{titlepage}
\renewcommand{\thefootnote}{\fnsymbol{footnote}}

\rightline{DAMTP-R94/11}
\vspace{0.1in}
\LARGE
\center{Generalised Point Interactions for the \\ Radial
Schr\"{o}dinger Equation\\ via Unitary Dilations}
\Large
\vspace{0.2in}
\center{C.J. Fewster\footnote{E-mail address:
C.J.Fewster@amtp.cam.ac.uk}}
\vspace{0.2in}
\large
\center{\em Department of Applied Mathematics and Theoretical Physics,
\\ University of Cambridge,
\\  Silver Street, Cambridge CB3 9EW, U.K.}
\vspace{0.2in}
\center{July 8, 1994}
\small\center{Revised November 28, 1994}
\vspace{0.2in}

\begin{abstract}
We present an inverse scattering construction of generalised point
interactions (GPI) -- point-like objects with non-trivial scattering
behaviour. The construction is developed for single centre $S$-wave
GPI models with rational $S$-matrices, and starts from an integral
transform suggested by the scattering data. The theory of unitary
dilations is then applied to construct a unitary mapping between
Pontryagin spaces which extend the usual position and momentum
Hilbert spaces. The GPI Hamiltonian is defined as a multiplication
operator on the momentum Pontryagin space and its free parameters are
fixed by a physical locality requirement. We determine the spectral
properties and domain of the Hamiltonian in general, and construct
the resolvent and M{\o}ller wave operators thus verifying that the
Hamiltonian exhibits the required scattering behaviour. The physical
Hilbert space is identified. The construction is illustrated by GPI
models representing the effective range approximation. For negative
effective range we recover a known class of GPI models, whilst the
positive effective range models appear to be new. We discuss the
interpretation of these models, along with possible extensions to our
construction.

\vspace{0.3truecm}
\end{abstract}

\setcounter{footnote}{0}
\renewcommand{\thefootnote}{\arabic{footnote}}
\end{titlepage}

\sect{Introduction and Main Ideas}

Generalised point interactions (GPI) are solvable models in quantum
mechanics representing point objects with non-trivial scattering
behaviour. The prototype for such models is the class of point
interactions (PI), corresponding to Hamiltonians with
$\delta$-function potentials essentially introduced by Fermi
\cite{Fermi} and rigorously defined as self-adjoint extensions of
$-\triangle$ on the domain of smooth functions compactly supported
away from the interaction centre \cite{BF}. This construction leads
to 1-parameter families of PI Hamiltonians in dimensions 2 and 3
which provide the leading order (scattering length) approximation to
the scattering behaviour of Schr\"{o}dinger operators with short
range potentials in the sector of zero angular momentum (see, for
example \cite{KF}). We refer to \cite{Alb} for an extensive
bibliography on PI models.

GPI models are employed to treat more general scattering behaviour,
such as higher order corrections to $S$-wave scattering, non-trivial
scattering in non-zero angular momentum sectors or point objects in
dimensions $d\ge 4$. Such models have  been studied for a long time
from the pseudo-potential viewpoint in many body physics -- see
\cite{Huang,GWu}. The mathematical study of such generalised point
interaction (GPI) models began in the Russian literature with the
work of Shirokov \cite{Shiro}, and more rigorous formulations were
later developed by Pavlov \cite{Pav1,Pav2,Pav3} and Shondin
\cite{Shond1,Shond2} (see also \cite{Diejen}). In contrast to PI
models, GPI Hamiltonians are not defined on the usual Hilbert space
$L^2(\RR^d)$, but on an extended space whose inner product may be
indefinite, in which case one must identify a physical Hilbert space
of states in order to recover the probability interpretation of
quantum mechanics.

In this paper, we introduce and develop a new inverse scattering
construction for single centre GPI models with non-trivial $S$-wave
scattering. It is currently an open problem to extend this
construction to higher angular momenta and dimensions $d\ge 4$; we
discuss this further in the Conclusion. Our method is based on the
technique of {\em unitary dilations} due in origin to Sz.-Nagy
\cite{Nagy} and extended by Davis \cite{Davis}.

The $S$-wave inverse scattering problem has been partially studied by
Shondin \cite{Shond1}, as we describe below. First, let us briefly
mention the two principal non-inverse GPI constructions, which we
call the {\em auxiliary space} and {\em distributional} methods. In
the auxiliary space method, developed by Pavlov and co-workers
\cite{Pav2,Pav3}, one starts with a given extended Hilbert space and
then seeks the class of GPI Hamiltonians which `live' on this space.
On the other hand, the distributional method of Shondin \cite{Shond2}
is a direct attempt to define $H=-\triangle+\ket{\omega}\bra{\omega}$,
where $\omega$ is somewhere in the $j<0$ portion of the scale of
Sobolev spaces $\HH_j=(-\triangle+1)^{-j/2}L^2(\RR^d)$. The
construction leads to a Pontryagin space\footnote{A Pontryagin space
is an indefinite (Krein) inner product space with a finite rank of
indefiniteness -- see Section~2.1.} $\Pi_m=\HH_0\oplus\CC^{2m}$,
where $m$ is the unique integer such that
$\omega\in\HH_{-m-2}\backslash\HH_{-m-1}$ and the inner product has
signature $(m,m)$ on the finite dimensional part. The GPI Hamiltonian
is then defined on $\Pi_m$ using Krein's formula \cite{Akh}.

Returning to the inverse problem, Shondin \cite{Shond1} considered
the inverse scattering problem in $d=3$ for $S$-wave scattering data
of form
\begin{equation}
\cot\delta_0(k) = k^{-1}r(k^2), \label{eq:GPIlo}
\end{equation}
where $r(z)$ is a rational function with real coefficients. Shondin
refers to this class of data as the `$R$ class': it corresponds
exactly with the class of rational $S$-wave $S$-matrices
satisfying the usual analytic continuation property $S(k)=S^*(-k)$
\cite{Newt}. In particular, the $R$ class contains truncated low
energy expansions of form
\begin{equation}
\cot\delta_0(k) = -\frac{1}{kL} + \frac{1}{2}kr_0 + k^3r_1+\cdots
+k^{2n+1}r_n, \label{eq:low}
\end{equation}
for low energy parameters $L,r_0,r_1,\ldots,r_n$ and any $n\ge 0$,
and thus furnishes approximations to the low energy $S$-wave
behaviour of Schr\"{o}dinger operators with short range potentials to
arbitrary order. Shondin's method starts by writing down a candidate
resolvent on a (positive definite) extension of the usual Hilbert
space. (In this respect it resembles the auxiliary space method).
Various free functions in this resolvent are then fixed by requiring
that the candidate be the resolvent of a self-adjoint operator.
However, this method is limited to those $r(z)$ with negative
imaginary part in the upper half plane, which is a somewhat
restrictive sub-class: for example, scattering data of
form~(\ref{eq:low}) is possible only with $r_1,\ldots ,r_n=0$ and
$r_0<0$. As we will see later, in the context of our method, more
general scattering data corresponds to GPI models defined on
Pontryagin spaces. Thus, in order to apply Shondin's method to such
data, one would have to guess not only the appropriate extension to
the Hilbert space, but also its inner product, thereby rendering it
much less practical as a construction.

In contrast, the method proposed here allows one to treat the full
$R$ class. It proceeds from the simple observation that, for
point-like interactions, the scattering data $\delta_0(k)$
completely specify the $S$-wave continuum eigenfunctions as $u_k(r)
= (2/\pi)^{1/2}\sin (kr+\delta_0(k))$, when scattering normalisation
is imposed. This is quite different from the usual situation in
inverse scattering theory, where the scattering data specifies (in
the first instance) only the asymptotic form of the $u_k(r)$. As a
result, the inverse scattering problem for GPI models can be solved
without recourse to the usual Gel'fand-Levitan machinery.

The $u_k(r)$ define an integral transform $\Tt$ between
$\HH_r=L^2((0,\infty),dr)$ and $\HH_k=L^2((0,\infty),dk)$, (i.e., the
radial position and momentum Hilbert spaces) by
\begin{equation}
(\Tt \psi)(k) =
\left(\frac{2}{\pi}\right)^{1/2}\int_0^\infty \sin
(kr+\delta_0(k))\psi(r) dr,
\end{equation}
We adopt $\Tt$ as a candidate for an eigenfunction transform
associated with our desired GPI Hamiltonian $\hg$. Of course, $\Tt$
cannot be the full eigenfunction transform unless it is unitary, in
which case we can define $\hg=\Tt^* k^2\Tt$. However, when $\Tt$ is
non-unitary, we may `dilate' it to a unitary operator between
enlarged inner product spaces, using the theory of unitary dilations
as follows. Firstly, we quantify the departure of $\Tt$ from
unitarity by the operators  $M_1=\II-\Tt\Tt^*$ and
$M_2=\II-\Tt^*\Tt$, and the closures $\MM_1$ and $\MM_2$ of their
respective ranges. For scattering data in the $R$ class, we will show
that the rank and signature of $M_1$ and $M_2$ are finite and given
in terms of indices reminiscent of Levinson's theorem \cite{Newt}.

Next, we define indefinite inner products on the $\MM_i$, together
with an operator
$\hat{\Tt}:\HH_r\oplus\MM_1\rightarrow\HH_k\oplus\MM_2$ which is
unitary with respect to the relevant inner products and satisfies
$P_{\HH_k}\hat{\Tt}|_{\HH_r}=\Tt$, where $P_{\HH_k}$ is the
orthoprojector onto $\HH_k$. $\hat{\Tt}$ is said to be a unitary
dilation\footnote{See Section~2 for a note on the nomenclature.}  of
$\Tt$. We emphasise that the construction of $\hat{\Tt}$ and the
enlarged inner product spaces requires no information beyond that
encoded in $\Tt$, and is unique up to a unitary equivalence and
further dilation. To complete the construction, we define the GPI
Hamiltonian $\hg$ by
\begin{equation}
\hg = \hat{\Tt}^\dagger
\left(\begin{array}{cc} k^2 & 0 \\ 0 & \Lambda \end{array}
\right)
\hat{\Tt},
\end{equation}
where the dagger denotes the Pontryagin space adjoint, and we have
used an obvious block matrix notation.  We will show that $\Lambda$
is completely determined by imposing a physical locality requirement:
that the `interaction' be localised at the origin. Mathematically,
this is expressed by requiring the Hamiltonian agrees with the free
Hamiltonian away from the interaction centre, i.e., $\hg(\psi,0)^T=
(-\psi^{\prime\prime},0)^T$ if $\psi\in C_0^\infty(0,\infty)$.
{}Subject to this locality condition, we have thus constructed both
$\hg$ and its spectral representation. The non-uniqueness in our
construction leads to a family of unitarily equivalent GPI
Hamiltonians with the same spectral and scattering properties.

Our plan is as follows. In Section~2, we briefly describe some
features of analysis in indefinite inner product spaces, and
also describe the construction of unitary dilations, essentially
following Davis \cite{Davis}. In addition, we sketch our construction
in a more abstract setting. Next, in Section~3, we explicitly
construct the operators $M_1$ and $M_2$ (which are of finite rank)
for a generic subclass of the $R$ class -- those whose scattering
amplitudes exhibit only simple poles on the physical sheet -- and
compute their rank and signature. In Section~4, we construct $\hg$ as
described above. Subject to the locality condition, we show that the
eigenvalues of $\hg$ occur at precisely those energies for which the
scattering amplitude derived from~(\ref{eq:GPIlo}) exhibits poles on
the physical sheet, as is the case for ordinary scattering from `nice'
potentials. We construct the corresponding eigenfunctions of $\hg$,
and isolate the physical Hilbert space. We also determine the domain
and resolvent of $\hg$, and explicitly construct its M{\o}ller wave
operators in a two-space setting, verifying that they exhibit the
required scattering theory.

In Section 5, we illustrate our procedure by constructing GPI models
with scattering behaviour $\cot\delta_0(k)=-1/(kL)+kM$, representing
the effective range approximation of low energy scattering theory
\cite{Newt}. In the case $M<0$, $\HH_r$ and $\HH_k$ are extended to
larger Hilbert spaces, and we recover the models of `type $B_2$'
previously constructed by Shondin \cite{Shond1}. These models also
arise as a special case of the auxiliary space construction in
\cite{Pav1}. The case $M>0$, for which Pontryagin spaces are
required, appears to be new. Our methods allow the entire  class of
GPI Hamiltonians to be constructed, along with their spectral
representations. A particularly interesting subclass of the models
constructed corresponds to the case $L=\infty$, with scattering theory
$\cot\delta_0(k)=kM$. Such models reproduce the leading order
behaviour of non-point interactions exhibiting a zero energy
resonance. We refer to these models as {\em resonance point
interactions} (RPI).

We also discuss how these GPI models may be used as models for
Schr\"{o}dinger operators with spherically symmetric potentials of
compact support. To do this, we employ a general methodology for
discussing the `large scale effects of small objects' developed by
Kay and the author \cite{KF}. In particular, we develop {\em fitting
formulae} (analogous to those given in \cite{KF}) for matching a
given potential $V(r)$ to the `best fit' GPI model. Finally, in
Section 6, we conclude by discussing various extensions to our
method.

The motivation for the present work arose in a consideration of the
scattering of charged particles off magnetic flux tubes of small
radius \cite{FK}, in which it was found that the scattering lengths
for spin-$\frac{1}{2}$ particles generically take the values $0$ or
$\infty$ in certain angular momentum sectors. In consequence, the
analogue of PI models representing dynamics in
the background of an infinitesimally thin wire of flux fails to
describe the leading order scattering theory in these sectors, and
should be replaced by models analogous to the RPI models mentioned
above. The special nature of this system can be attributed to the
fact that it is an example of supersymmetric quantum mechanics.
Elsewhere \cite{F}, we will construct the appropriate class of RPI
for this system.

\sect{Preliminaries}
\subsection{Unitary Dilations}

We begin by describing the unitary dilation theory required in the
sequel. Let $\HH_1,\ldots,\HH_4$ be Hilbert spaces and
$T\in\LL(\HH_1,\HH_2)$. Then
$\hat{T}\in\LL(\HH_1\oplus\HH_3,\HH_2\oplus\HH_4)$ is called a
{\em dilation} of $T$ if
$T= P_{\HH_2}\hat{T}|_{\HH_1}$
where $P_{\HH_2}$ is the orthogonal projector onto $\HH_2$. In
block matrix form, $\hat{T}$ takes form
\begin{equation}
\hat{T} = \left(\begin{array}{cc} T & P \\ Q & R \end{array}\right).
\label{eq:dilfm}
\end{equation}
Our nomenclature follows that of Halmos \cite{Halmos}. Elsewhere
(e.g., in the work of Davis \cite{Davis}), the term `dilation' (or
`dilatation') often means that $\hat{T}^n$ is a dilation of $T^n$
and $(\hat{T}^*)^n$ is a dilation of $(T^*)^n$ for each $n=1,2,\ldots$
(in addition, $\HH_1=\HH_2$, and $\HH_3=\HH_4$). We refer to such
operators as {\em power dilations}: in the block
form~(\ref{eq:dilfm}), this requires $PR^nQ=0$ for each
$n=0,1,2,\ldots$.

According to a result of Sz.-Nagy \cite{Nagy}, any contraction $T$
from one Hilbert space to another (i.e., a bounded operator
satisfying $\|T\|\le 1$) has a unitary dilation between larger
Hilbert spaces. Subsequently, Davis \cite{Davis} extended this
result to arbitrary closed densely defined operators at the
cost of introducing indefinite inner product spaces. (It is clear
that if $\|T\|>1$, no Hilbert space unitary dilation is possible.)
In fact, Davis' construction yields a unitary {\em power} dilation
of the original operator.  This has no physical relevance in our
construction, and so we use a more economical `cut-down' version of
Davis' result, described below. First, we briefly review the salient
features of analysis in indefinite inner product spaces. Full
treatments can be found in the monographs of Bogn\'ar \cite{Bognar}
and Azizov and Iokhvidov~\cite{Azizov}.

We employ a particular class of indefinite inner product
spaces known as {\em $J$-spaces}. Let $\HH$ be a Hilbert space with
(positive definite) inner product $\inner{\cdot}{\cdot}$,
equipped with a unitary involution, $J$. We define a non-degenerate
indefinite inner product $[\cdot,\cdot]$ on $\HH$ by
\begin{equation}
[x,y]=\inner{x}{Jy},
\end{equation}
which we call the {\em $J$-inner product}. $\HH$ equipped with the
$J$-inner product is called a $J$-space. $\HH$ admits decomposition
$\HH=\HH_+\oplus\HH_-=\HH_+[+]\HH_-$ into the eigenspaces $\HH_\pm$
of $J$ with eigenvalue $\pm 1$, where $[+]$ denotes the orthogonal
direct sum in the $J$-inner product. If at least one of the
$\HH_\pm$ is finite dimensional, then $\HH$ is a {\em Pontryagin
space} with respect to $[\cdot,\cdot]$ .

The topology of a $J$-space is determined by the Hilbert space norm;
however, operator adjoints and the notion of unitarity are defined
relative to the $J$-inner product. Thus if $\HH_i$ ($i=1,2$) are
$J_i$-spaces, and $T\in\LL(\HH_1,\HH_2)$, the {\em
$(J_1,J_2)$-adjoint} $T^\dagger$ of $T$ is defined in terms of the
Hilbert space adjoint $T^*$ by
\begin{equation}
T^\dagger = J_1T^*J_2.
\end{equation}
Equivalently, $[T^\dagger x,y]_{\HH_1}=[x,Ty]_{\HH_2}$ for all
$x\in\HH_2$, $y\in\HH_1$. If $[Ux,Uy]_{\HH_2}=[x,y]_{\HH_1}$ for all
$x,y\in \DD\subset\HH_1$, $U$ is said to be {\em
$(J_1,J_2)$-isometric}; if in addition $U$ is a linear isomorphism of
$\HH_1$ and $\HH_2$, and $\DD=\HH_1$, $U$ is said to be {\em
$(J_1,J_2)$-unitary}. Equivalently, $UU^\dagger = \II_{\HH_1}$ and
$U^\dagger U = \II_{\HH_2}$. If $\HH_1=\HH_2$ with $J_1=J_2=J$, terms
such as $(J_1,J_2)$-isometric are abbreviated to $J$-isometric etc.

Returning to the construction of unitary dilations, let $T$ be any
bounded operator $T\in\LL(\HH_1,\HH_2)$, and define operators
$M_1 = \II-TT^*$ and $M_2=\II-T^*T$. It is trivial to show that the
respective closures $\MM_i=\overline{\Ran M_i}$ of their ranges are
$\sgn (M_i)$-spaces, and hence that $\KK_i=\HH_i\oplus\MM_i$ are
$J_i$-spaces, where $J_i=\II_{\HH_i}\oplus\sgn(M_i)$. We now define
a dilation $\hat{T}$ of $T$ by
\begin{equation}
\hat{T} = \left(\begin{array}{cc} T & -\sgn(M_1)|M_1|^{1/2} \\
                                  |M_2|^{1/2} & T^*|_{\MM_1}
                \end{array} \right),
\end{equation}
which has $(J_1,J_2)$-adjoint $\hat{T}^\dagger$ equal to
\begin{equation}
\hat{T}^\dagger = J_1\hat{T}^*J_2=
 \left(\begin{array}{cc}
                            T^* & \sgn (M_2) |M_2|^{1/2} \\
                            - |M_1|^{1/2} & T|_{\MM_2}
                        \end{array} \right).
\end{equation}
Here, we have used the intertwining relations $Tf(T^*T)=f(TT^*)T$ and
$T^*f(TT^*)=f(T^*T)T^*$, which hold for any continuous Borel function
$f$. It is now easy to show that $\hat{T}^\dagger\hat{T}=\II_{\KK_1}$
and $\hat{T}\hat{T}^\dagger=\II_{\KK_2}$, thus verifying that
$\hat{T}$ is a $(J_1,J_2)$-unitary dilation of $T$. In our
application, $M_1$ and $M_2$ are finite rank, and so the $J$-spaces
constructed above are Pontryagin spaces.

We briefly consider the uniqueness of the unitary dilations
constructed above. Suppose $\NNN_i$ are $J_i$
spaces ($i=1,2$) and that $\tilde{T}:\HH_1\oplus\NNN_1 \rightarrow
\HH_2\oplus\NNN_2$ is a unitary dilation of $T$ with matrix
form~(\ref{eq:dilfm}). Then, provided that the $M_i$ are finite
rank, one may show that
\begin{equation}
P_{\HH_2\oplus \Qt}
\tilde{T}|_{\HH_1\oplus \Pt}=
\left(\begin{array}{cc} \II & 0 \\ 0 & U_2 \end{array}\right)
\hat{T}
\left(\begin{array}{cc} \II & 0 \\ 0 & U_1^\dagger
\end{array}\right),
\end{equation}
where $\Pt=P^\dagger\overline{\Ran M_1}$, $\Qt=Q\overline{\Ran
M_1}$, and $U_1$ and $U_2$ are unitaries (with respect to the
$J$-inner products) from $\MM_1$ and $\MM_2$ to $\Pt$ and $\Qt$
respectively. In addition, $P_{\HH_2\oplus \Qt}$ is an orthogonal
projection onto $\HH_2\oplus\Qt$ in $\HH_2\oplus\NNN_2$.

Thus $\hat{T}$ is unique up to further dilation and unitary
equivalence of the above form. If the $M_i$ are not of finite rank,
this statement also holds if the $M_i$ are strictly positive. More
generally, it is not clear whether $\Qt$ is necessarily
orthocomplemented, and therefore whether $P_{\HH_2\oplus \Qt}$ exists.

\subsection{Abstract Setting}
\label{sect:abs}

In this section, we sketch our construction in a general setting,
which makes clear how it may be extended. In particular, we show how
the domain and action of the Hamiltonian is determined.

Let $\HH_i$ ($i=1,2$) be Hilbert spaces and let $A$ be a densely
defined symmetric operator with domain $\DD\subset \HH_1$. Suppose
that $A$ possesses two self-adjoint extensions $A_\pm$ such that
\begin{equation}
A_\pm = \Tt_\pm^* \tilde{A}\Tt_\pm
\end{equation}
where $\tilde{A}$ is a self-adjoint operator on $\HH_2$ with
$(\tilde{A}+\omega)^{-1}$ bounded for some $\omega\in\RR$, and
$\Tt_\pm$ are unitary operators $\Tt_\pm :\HH_1\rightarrow\HH_2$. Let
$a_+$ and $a_-$ be bounded operators on $\HH_2$ which commute with
$\tilde{A}$ and define
\begin{equation}
\Tt = a_+\Tt_+ + a_-\Tt_-.
\end{equation}
In our application, $a_\pm$ are determined by the scattering
data. We define $M_1$ and $M_2$ as above, for simplicity
assuming that they are finite rank (as they are in our application).
The unitary dilation $\hat{\Tt}$ derived above is then used to
define a self-adjoint operator $B$ on the Pontryagin space $\Pi_1
=\HH_1\oplus\MM_1$ by
\begin{equation}
B = \hat{\Tt}^\dagger
\left(\begin{array}{cc} \tilde{A} & 0 \\ 0 & \Lambda \end{array}
\right)
\hat{\Tt},
\end{equation}
where $\Lambda$ is a self-adjoint operator on $\MM_2$ (with respect
to its inner product). Thus
\begin{equation}
B\left(\begin{array}{c} \varphi\\ \Phi\end{array}\right)=
\left(\begin{array}{c}
\Tt^*\tilde{A}(\Tt\varphi-\Theta)
+\sgn M_2|M_2|^{1/2}\Lambda(|M_2|^{1/2}\varphi+\Tt^*|_{\MM_1}\Phi) \\
-|M_1|^{1/2}\tilde{A}(\Tt\varphi-\Theta)
+\Tt|_{\MM_2}\Lambda (|M_2|^{1/2}\varphi+\Tt^*|_{\MM_1}\Phi)
\end{array}\right),
\label{eq:actB}
\end{equation}
where $\Theta=\sgn M_1 |M_1|^{1/2}\Phi$ (considered as an element of
$\HH_2$), and $B$ has domain
\begin{equation}
D(B) = \{ (\varphi,\Phi)^T\mid \Tt\varphi-\Theta
\in D(\tilde{A})\}. \label{eq:domB}
\end{equation}
{}To gain a more explicit description of $D(B)$, we impose the
requirement that $B$ be a self-adjoint extension of the
{\em non-densely defined} operator $A\oplus 0$ on $\DD\oplus
0\subset\Pi_1$, i.e., $B(\varphi,0)^T=(A\varphi,0)^T$ for all
$\varphi\in\DD$. Later this will carry the physical interpretation of
a locality condition. It is easy to show that this requirement is
satisfied if and only if $\MM_2$ is invariant under $A^*$ and
\begin{equation}
\Lambda=(|M_2|^{-1/2}A^*|_{\MM_2}|M_2|^{1/2})^*.
\end{equation}

As a consequence of locality, we note that if  $(\varphi,\Phi)^T\in
D(B)$ with $B (\varphi,\Phi)^T=(\tilde{\varphi},\tilde{\Phi})^T$,
then  $\varphi\in D(A^*)$, and $\tilde{\varphi}=A^*\varphi$. For take
any $\psi\in \DD$. Then
\begin{equation}
\inner{\tilde{\varphi}}{\psi}_{\HH_1} =
\left[
\left(\begin{array}{c}\tilde{\varphi}\\
\tilde{\Phi}\end{array}\right), \left(\begin{array}{c}\psi\\
0\end{array}\right)\right]_{\Pi_1} =
\left[
\left(\begin{array}{c}\varphi\\ \Phi\end{array}\right),
B\left(\begin{array}{c}\psi\\ 0\end{array}\right)
\right]_{\Pi_1}= \inner{\varphi}{A\psi}_{\HH_1}.
\end{equation}
We may therefore re-write~(\ref{eq:domB}) as
\begin{equation}
D(B)=\left\{ \left(\begin{array}{c} \varphi \\ \Phi \end{array}
\right) \mid \varphi\in D(A^*), \quad \Theta_1 \in D(\tilde{A})
\right\},
\end{equation}
where $\Theta_1 =a_+\Tt_+\chi_+ + a_-\Tt_-\chi_-+\Theta$ and
$\chi_\pm=(A_\pm+\omega)^{-1}(A^*+\omega)\varphi-\varphi$. The
advantage of this expression is that $\chi_\pm$ can be shown to be the
unique element of $\ker (A^*+\omega)$ such that $\varphi+\chi_\pm\in
D(A_\pm)$. In our application, $\chi_\pm$ may be expressed in terms
of the value of $\varphi$ and its first derivative at the origin.

{}To determine the action of $B$ more explicitly, we use the fact that
the upper component of the right-hand side of~(\ref{eq:actB}) is
equal to $A^*\varphi$ in order to compute $\tilde{\Theta}=\sgn
M_1|M_1|^{1/2} \tilde{\Phi}$. We obtain
\begin{equation}
\tilde{\Theta}=-M_1\tilde{A}(\Tt\varphi-\Theta)+
\Tt(A^*\varphi-\Tt^*\tilde{A}(\Tt\varphi-\Theta))=
\Tt A^*\varphi-\tilde{A}(\Tt\varphi-\Theta)
\end{equation}
Using the fact that $\Theta_1\in D(\tilde{A})$, this becomes
\begin{equation}
\tilde{\Theta}=\tilde{A}\Theta_1 +\omega(\Theta_1-\Theta) +
\Tt(A^*+\omega)\varphi-(\tilde{A}+\omega)
(\Tt\varphi+\Theta_1-\Theta).
\end{equation}
The last two terms cancel by definition of $\chi_\pm$ and we
conclude that
\begin{equation}
B\left(\begin{array}{c}\varphi\\ \Phi\end{array}\right)=
\left(\begin{array}{c} A^*\varphi \\
(\sgn M_1 |M_1|^{1/2})^{-1}\tilde{\Theta}
\end{array}\right)
\end{equation}
where $\tilde{\Theta}=\tilde{A}\Theta_1 + \omega(a_+\Tt_+\chi_+ +
a_-\Tt_-\chi_-)$.

\sect{Determination of $M_1$ and $M_2$}

In this section, we determine the operators $M_1=\II-\Tt\Tt^*$ and
$M_2=\II-\Tt^*\Tt$, where $\Tt$ is an integral transformation
arising from the scattering data in the Shondin $R$ class
\cite{Shond1} given by
\begin{equation}
\cot\delta_0(k) = k^{-1}\frac{p(k^2)}{q(k^2)},\qquad
\delta_\ell(k)\equiv 0\quad {\rm for}~\ell\ge 1,
\label{eq:GPIlow}
\end{equation}
where $p(z)$ and $q(z)$ are coprime polynomials in $\RR[z]$, the
ring of polynomials with real coefficients. In particular, we will
show how the rank and signature of the $M_i$ are determined by two
`Levinson indices' defined below. We emphasise that our methods are
very different to those of Shondin.

The scattering amplitude corresponding to $\delta_0(k)$ is
\begin{equation}
f_0(k) = \frac{1}{k}e^{i\delta_0(k)}\sin\delta_0(k)=
\frac{q(k^2)}{p(k^2) - ikq(k^2)}.
\end{equation}
Defining the polynomial $W(z)$ by
\begin{equation}
W(z) =\left\{\begin{array}{cl} p(-z^2)-zq(-z^2) & p(0)\not=0 \\
                               p(-z^2)/z-q(-z^2) & p(0)=0,
\end{array}\right. \label{eq:W}
\end{equation}
we note that $f_0(k)$ exhibits poles where $W(ik)=0$. The set
$\Omega$ of zeros of $W(z)$ in the left-hand half-plane ${\rm Re}\,
z<0$ corresponds to poles of $f_0(k)$ such that $k^2$ lies on the
physical sheet. We refer to the situation where these poles
(and hence the corresponding zeros of $W(z)$) are simple as the
{\em generic case}. In Theorem~\ref{Thm:local}, we will show
that the discrete spectrum of the GPI Hamiltonian is precisely
$\{E=-\omega^2\mid\omega\in\Omega\}$ under the requirement of
locality.\footnote{These eigenvalues can be complex: we will return
to this point in section~5.3.}

The qualitative features of the scattering data~(\ref{eq:GPIlow})
are described by the degrees of $p$ and $q$, two indices $I_L^\pm$
defined below, and the asymptotic behaviour of $\cot\delta_0(k)$
given by
\begin{equation}
\sigma_0 = \sgn\lim_{k\rightarrow 0^+}
\cot\delta_0(k)\qquad {\rm and}\qquad \sigma_\infty
=\sgn\lim_{k\rightarrow\infty}\cot\delta_0(k),
\end{equation}
where the limits are allowed to be $\pm\infty$. The indices
$I_L^\pm$ are defined by
\begin{equation}
I_L^+= \frac{\delta_0(0)-\delta_0(\infty)}{\pi}\qquad{\rm and}
\qquad
I_L^-=\frac{\zeta(0)-\zeta(\infty)}{\pi},
\end{equation}
where the auxiliary scattering data $\zeta(k)$ is defined
as a continuous function on $\RR^+$ by
\begin{equation}
\cot\zeta(k) = -k^{-1}\frac{p(-k^2)}{q(-k^2)}. \label{eq:zeta}
\end{equation}
We refer to $I_L^\pm$ as the Levinson indices (although Levinson's
theorem \cite{Newt} will not hold in its usual form).

We now define the integral transform
$\Tt=\cos\delta_0(k)\St+\sin\delta_0(k)\Ct$, which is suggested by
the
na\"{\i}ve generalised eigenfunctions $u_k(r) =
(2/\pi)^{1/2}\sin (kr+\delta_0(k))$.
Here, $\St$ and $\Ct$ are the sine and cosine transforms,
defined by
\begin{eqnarray}
(\St \psi)(k) =
\sqrt{\frac{2}{\pi}}\int_0^\infty dr\, \psi(r)\sin kr  & {\rm and}
& (\Ct \psi)(k) =
\sqrt{\frac{2}{\pi}}\int_0^\infty dr\, \psi(r)\cos kr
\end{eqnarray}
(the integrals are intended as limits in $L^2$-norm).
Both are unitary maps from $\HH_r$ to $\HH_k$;
their inverses have the same form, with $r$ and $k$ exchanged.
Thus $\Tt$ is given explicitly by
\begin{equation}
\Tt = \frac{p(k^2)}{(p(k^2)^2+k^2q(k^2)^2)^{1/2}}\St +
\frac{kq(k^2)}{(p(k^2)^2+k^2q(k^2)^2)^{1/2}} \Ct.
\end{equation}
Because $\St$ and $\Ct$ furnish the spectral representations of
$-d^2/dr^2$ on $L^2(\RR^+)$ with Dirichlet and Neumann boundary
conditions respectively at the origin, we are in the general situation
of Section~2.2.

We now restrict to the generic case and explicitly construct the
$M_i$ and compute their rank and signature. $M_2$ is given by the
following proposition, whose proof is given later in this section.
\begin{Prop}
\label{Prop:M2} In the generic case,
\begin{equation}
M_2 = \sum_{\omega\in\Omega} \alpha_\omega
\ket{\xi_\omega}\bra{\xi_{\overline{\omega}}}, \label{eq:M2}
\end{equation}
where $\xi_\omega(r) =e^{\omega r}$, and $\alpha_\omega$ is the
residue
\begin{equation}
\alpha_\omega= {\rm Res}_\omega
2zf_0(-iz). \label{eq:aw}
\end{equation}
In addition, $\Ran
M_2=\Span\{\xi_\omega\mid\omega\in\Omega\}$,
and
\begin{eqnarray}
\rank M_2 &=& \frac{1}{2}\deg W + I_L^+  \label{eq:M2rnk} \\
\sig M_2 &=& \frac{1}{2}\left(\sigma_0^2-\sigma_\infty^2\right)
-I_L^- . \label{eq:M2sig}
\end{eqnarray}
\end{Prop}

Next, define $\MM_1$ to be the space of all
$L^2$-vectors of form $Q(k^2)k(p(k^2)^2+k^2q(k^2)^2)^{-1/2}$,
such that $Q(z)\in \CC[z]$ is a polynomial with complex
coefficients. Thus
\begin{equation}
\MM_1 =
(p(k^2)^2+k^2q(k^2)^2)^{-1/2}k\CC_{\mho-1}[k^2]
\end{equation}
where $\CC_r[z]$ is the $r+1$-dimensional complex vector space of
polynomials with complex coefficients and degree at most $r$,
and $\mho=\dim \MM_1$ is given by
\begin{equation}
\mho=\frac{1}{2}\deg W+\frac{1}{2}(\sigma_\infty^2-\sigma_0^2) =
\max\{\deg p ,\deg q\}.
\end{equation}
$M_1$ is described by
\begin{Prop} \label{Prop:M1}
In the generic case, $M_1$ vanishes on
${\MM_1}^\perp$, and its action on $\MM_1$ is given by
$M_1 Q(k^2)k(p(k^2)^2+k^2q(k^2)^2)^{-1/2}=\tilde{Q}(k^2)k
(p(k^2)^2+k^2q(k^2)^2)^{-1/2}$, where
\begin{equation}
\tilde{Q}(k^2)=
Q(k^2)+\sum_{\omega\in\Omega}
\frac{Q(-\omega^2)\alpha_\omega}{q(-\omega^2)}
\frac{p(k^2)-\omega q(k^2)}{\omega^2+k^2}.
\end{equation}
Moreover, $\Ran M_1=\MM_1$ and
\begin{eqnarray}
\rank M_1 & = &\frac{1}{2}\deg W +
\frac{1}{2}(\sigma_\infty^2-\sigma_0^2) \label{eq:M1rnk} \\
\sig M_1 &=& -(I_L^+ +I_L^-). \label{eq:M1sig}
\end{eqnarray}
\end{Prop}

As an example, let us consider the sub-class of the $R$ class
considered by Shondin \cite{Shond1}; namely, the case where
$r(z)=p(z)/q(z)$ has negative imaginary part in the upper half-plane.
In this case, it is easy to show that there can be no solutions to
$r(-z^2)=z$ and hence to $W(z)=0$ in the left-hand half-plane,
except on the real axis. Moreover, one can show that the residues
$\alpha_\omega$ at these zeros are necessarily positive, so $M_2$ is
a positive operator as a result of~(\ref{eq:M2}). Accordingly, $\Tt$
is contractive, and our method yields a unitary dilation defined on
Hilbert spaces. This explains why Shondin was able to construct
these GPI models on enlarged {\em Hilbert} spaces.

We now prove the above propositions.

{\noindent\em Proof of Proposition~\ref{Prop:M2}:}
$M_2$ may be written in two equivalent forms:
\begin{eqnarray}
M_2 &=& \St^{-1}\sin^2\delta_0(k)\St
 -\Ct^{-1}\sin^2\delta_0(k)\Ct   \nonumber \\
& &-\Ct^{-1}\sin\delta_0(k)\cos\delta_0(k)\St -
\St^{-1}\sin\delta_0(k)\cos\delta_0(k)\Ct \label{eq:ker1}\\
&=& \Ct^{-1}\cos^2\delta_0(k)\Ct  -\St^{-1}\cos^2\delta_0(k)\St
\nonumber \\
& &-\Ct^{-1}\sin\delta_0(k)\cos\delta_0(k)\St -
\St^{-1}\sin\delta_0(k)\cos\delta_0(k)\Ct . \label{eq:ker2}
\end{eqnarray}
{}To convert this into an integral kernel we use the following Lemma,
which may be proved by standard means (cf. Theorem IX.29 in
\cite{RSii}). Here, $v(x)$ and $w(x)$ stand for either $\sin x$ or
$\cos x$, and $\Vt$ and $\Wt$ are the corresponding integral
transforms from $\HH_r$ to $\HH_k$.
\begin{Lem} \label{Lem:ker}
Let $g(k)\in L^2(\RR^+)\cap L^\infty(\RR^+)$ and define
$G=\Vt^{-1}g(k)\Wt$. Then $G$ has integral kernel
\begin{equation}
G(r,r^\prime) = \frac{2}{\pi} \int_0^\infty v(kr)w(kr^\prime)g(k)dk,
\end{equation}
(where the integral is a limit in $L^2$-norm).
\end{Lem}

In the case $\deg p>\deg q$, $\sin^2\delta_0(k)$ and
$\sin\delta_0(k)\cos\delta_0(k)$ are $L^2\cap L^\infty$ and so,
applying Lemma~\ref{Lem:ker} to~(\ref{eq:ker1}) and combining
terms, $M_2$ has integral kernel
\begin{equation}
M_2(r,r^\prime) =
\frac{i}{\pi}\int_{-\infty}^\infty e^{i\delta_0(k)}
\sin\delta_0(k) e^{ik(r+r^\prime)}dk
=\frac{1}{\pi}\int_{-\infty}^\infty
\frac{ikq(k^2)e^{ik(r+r^\prime)}}{p(k^2)-ikq(k^2)} dk.
\end{equation}
Making the substitution $z=ik$ and closing the contour in the
left-hand half-plane, the integrand has a simple pole at each
$\omega\in\Omega$ and~(\ref{eq:M2}) follows. If $\deg q\ge\deg p$, we
argue similarly using~(\ref{eq:ker2}) to obtain the same result as
before.

By linear independence of the $\xi_\omega$ and non-vanishing of the
$\alpha_\omega$, it follows that $\Ran
M_2=\MM_2=\Span\{\xi_\omega\mid\omega\in\Omega\}$, so $\rank
M_2=|\Omega|$, the cardinality of $\Omega$.
Using residue calculus, one may show that
\begin{equation}
|\Omega| = \frac{1}{2}\deg W +
\frac{1}{2\pi}\int_{-\infty}^\infty \frac{W^\prime(ik)}{W(ik)} dk.
\end{equation}
By rewriting the second term as an integral over $(0,\infty)$, a
small amount of algebra shows that the integrand is
$-\pi^{-1}\delta^\prime_0(k)$. Thus~(\ref{eq:M2rnk}) is established.

{}To compute $\sig M_2$, we define the hermitian
form $m_2(\varphi,\psi): \MM_2\times\MM_2\rightarrow \CC$ by
$m_2(\varphi,\psi)=\inner{\varphi}{M_2\psi}$. Labelling the
elements of $\Omega$ as $\omega_1,\ldots,\omega_{|\Omega|}$, and
writing $\psi=\sum_i c_i\xi_{\omega_i}$, we have
\begin{equation}
m_2(\psi,\psi) = \sum_{i,j,k} \overline{c_i}
\inner{\xi_{\omega_i}}{\xi_{\omega_j}}\alpha_j
\inner{\xi_{\overline{\omega_j}}}{\xi_{\omega_k}}c_k
= c^\dagger\Xi^\dagger A\Xi c,
\end{equation}
where $A$ and $\Xi$ are hermitian. $\Xi$ has components
$\Xi_{ij}=\inner{\xi_{\omega_i}}{\xi_{\omega_j}}$,
and is non-singular by
linear independence of the $\xi_\omega$. By Sylvester's Law of
Inertia \cite{Cohn}, the signature of $M_2$ equals that of $A$,
which has components
\begin{equation}
A_{ij} = \left\{ \begin{array}{cl}
\alpha_{\omega_i} & \omega_i=\overline{\omega_j} \\
0 & {\rm otherwise}. \end{array}\right.
\end{equation}
$A$ has eigenvalues $\{\alpha_\omega\mid \omega\in\RR\}\cup \{\pm
|\alpha_\omega|\mid\omega\not\in\RR\}$.
Labelling the $\omega_i$ so that $\omega_1,\ldots,\omega_r$ are
the real elements of $\Omega$, we therefore have
$\sig M_2=\sig\diag (\alpha_{\omega_1}\ldots,\alpha_{\omega_r})$.
(We have used the fact that
$\alpha_{\overline{\omega}}=\overline{\alpha_\omega}$, and
in particular that $\omega_r\in\RR$ implies
$\alpha_r\in\RR$.) Defining $\zeta(k)$ by~(\ref{eq:zeta}), it is easy
to show that $\cot\zeta(-\omega)=1$ for $\omega\in\Omega$, and that
\begin{equation}
\alpha_\omega =2\lim_{z\rightarrow -\omega}
\frac{z+\omega}{1-\cot\zeta(z)} = \frac{1}{\zeta^\prime(-\omega)}.
\end{equation}
Thus $\sig\diag(\alpha_1,\ldots,\alpha_r)$ is equal to the number of
times that $\zeta(k)\equiv\pi/4 \pmod\pi$ as $k$
traverses $\RR^+$, counted according to the sign of
$\zeta^\prime(k)$ at such points. This is related to the Levinson
index $I_L^-$ by~(\ref{eq:M2sig}). $\square$

{\noindent\em Proof of Proposition~\ref{Prop:M1}:} We compute
\begin{eqnarray}
M_1 &=& -\frac{p(k^2)}{(p(k^2)^2+k^2q(k^2)^2)^{1/2}}\St\Ct^{-1}
\frac{kq(k^2)}{(p(k^2)^2+k^2q(k^2)^2)^{1/2}}  \nonumber \\
&& -\frac{kq(k^2)}{(p(k^2)^2+k^2q(k^2)^2)^{1/2}}\Ct\St^{-1}
\frac{p(k^2)}{(p(k^2)^2+k^2q(k^2)^2)^{1/2}},
\end{eqnarray}
which vanishes identically on the closure of
$\DD=(p(k^2)^2+k^2q(k^2))^{1/2}\St C_0^\infty(0,\infty)$ as a result
of elementary properties of the sine and cosine transforms.
Furthermore, $\overline{\DD}^\perp$ is precisely the space $\MM_1$
defined above, because $\psi\perp\DD$ if and only if
$(p(k^2)^2+k^2q(k^2))^{1/2}\psi$ is the sine transform of a
distribution supported at the origin and therefore an odd polynomial
(cf. Theorem~V.11 in \cite{RSi}). Hence $M_1$ vanishes on
$\MM_1^\perp$ and $\Ran M_1\subset \MM_1$.

Next, we compute the action of $M_1$ on $\MM_1$. By contour
integration,
\begin{equation}
\Tt^*\frac{kQ(k^2)}{(p(k^2)^2+k^2q(k^2)^2)^{1/2}}
=-\left(\frac{\pi}{2}\right)^{1/2}\sum_{\omega\in\Omega}
\frac{Q(-\omega^2)\alpha_\omega}{q(-\omega^2)}\xi_\omega(r),
\label{eq:Tstr}
\end{equation}
for polynomials $Q(z)$ such that the operand is in $L^2$. Moreover,
it is easy to show that
\begin{equation}
T\xi_{\omega} = \left(\frac{2}{\pi}\right)^{1/2}
\frac{k}{(p(k^2)^2+k^2q(k^2)^2)^{1/2}}
\frac{p(k^2)-\omega q(k^2)}{\omega^2+k^2},
\label{eq:Txi}
\end{equation}
from which the action of $M_1$ can be read off as required.

{}To compute the rank and signature of $M_1$, we use the fact that
\begin{equation}
\rank M_1 - \rank M_2 = \sig M_1 -\sig M_2 =
\dim\ker \Tt^*-\dim\ker\Tt ,
\end{equation}
which follows from the intertwining relations $M_1\Tt=\Tt M_2$ and
$M_2\Tt^*=\Tt^*M_1$. It therefore remains to determine the dimensions
of the relevant kernels. Firstly, note that $\ker \Tt^*\subset\MM_1$
and that (from~(\ref{eq:Tstr}))
$\psi=Q(k^2)k(p(k^2)^2+k^2q(k^2)^2)^{-1/2}\in\ker \Tt^*$ if and only
if $\psi\in\MM_1$ and $Q(-\omega^2)=0$ for each $\omega\in\Omega$.
Thus $\prod_{\omega\in\Omega} (z+\omega^2)$ divides $Q(z)$ and so
\begin{equation}
\dim\ker\Tt^* = \min\{\mho-|\Omega|,0\}.
\end{equation}

Now consider $\ker\Tt$. We note that~(\ref{eq:Txi}) may be rewritten
\begin{equation}
q(-\omega_i^2)\Tt\xi_{\omega_i}=\left(\frac{2}{\pi}\right)^{1/2}
\frac{k}{(p(k^2)^2+k^2q(k^2)^2)^{1/2}}
\frac{p(k^2)q(-\omega_i^2)-p(-\omega_i^2)q(k^2)}{k^2+\omega_i^2},
\end{equation}
and apply the following abstract algebraic result:
\begin{Lem}
\label{Lem:FW}
Let $Q,R\in \CC[z]$ be coprime with $\max\{\deg Q,\deg R\}=k\ge 0$,
and let $\lambda_1,\ldots,\lambda_m$ be distinct elements of $\CC$.
Then the polynomials $P_1(z),\ldots P_m(z)$, defined by
\begin{equation}
(z-\lambda_i) P_i(z) = R(\lambda_i)Q(z)-Q(\lambda_i)R(z)
\end{equation}
span a $\min\{k,m\}$-dimensional subspace of $\CC_{k-1}[z]$.
\end{Lem}
{\em Proof:} Let  $n=\min\{k,m\}$. Then it is enough to show that
$P_1,\ldots,P_n$ are linearly independent. Assuming that $\deg Q=k$,
we note that  $P_i(z)=R(z) \tilde{Q}_i(z)-Q(z)\tilde{R}_i(z)$, where
$\tilde{Q}_i(z)=(Q(z)-Q(\lambda_i))/(z-\lambda_i)$ and
$\tilde{R}_i(z)=(R(z)-R(\lambda_i))/(z-\lambda_i)$. Suppose the $P_i$
are linearly dependent. Then $R(z) S(z) = Q(z) T(z)$ where $S(z)
=\sum_i \alpha_i \tilde{Q}_i(z)$ and $T(z) =\sum_i\alpha_i
\tilde{R}_i(z)$, for some $0\not= (\alpha_1,\ldots,\alpha_n)^T\in
\CC^n$. Because $Q$ and $R$ are coprime, this implies that $S$ and
$T$ vanish identically. But one may easily show that the
$\tilde{Q}_i$ are linearly independent, by explicitly considering
their coefficients. We therefore obtain a contradiction. $\square$

In our application, $m=|\Omega|$ with $\lambda_i=-\omega_i^2$ for
each $i=1,\ldots,m$ and $k=\max\{\deg p,\deg q\}=\mho$. Thus
$\dim\Tt\Ran M_2=\min\{|\Omega|,\mho\}$ and so
\begin{equation}
\dim\ker\Tt= \min\{|\Omega|-\mho,0\}.
\end{equation}
It follows that $\rank M_1-\rank M_2=\sig M_1-\sig
M_2=\mho-|\Omega|$, from which~(\ref{eq:M1rnk}) and~(\ref{eq:M1sig})
follow. $\square$

\sect{The GPI Hamiltonian}
\label{sect:GHm}
\subsection{Locality and Spectral Properties}

The results of the previous two sections allow the construction of a
unitary dilation $\hat{\Tt}$ of the integral transform $\Tt$. Here,
we employ $\hat{\Tt}$ to define a GPI Hamiltonian consistent with
scattering theory~(\ref{eq:GPIlow}). We denote
$\Pi_r=\HH_r\oplus\MM_1$ and $\Pi_k=\HH_k\oplus\MM_2$ with $J$-inner
products specified by $J_r=\II_{\HH_r}\oplus\sgn (M_1)$, and
$J_k=\II_{\HH_k}\oplus\sgn (M_2)$. In terms of our general
discussion in Section~2.2, we set $A=-d^2/dr^2$ on domain
$C_0^\infty(0,\infty)$, and define $\Tt_+=\St$, $\Tt_-=\Ct$, setting
$a_+$ and $a_-$ to be multiplication by $\cos\delta_0(k)$ and
$\sin\delta_0(k)$ respectively. Thus
$A_+=\St^*k^2\St$, the self-adjoint extension of $A$ with Dirichlet
boundary conditions at the origin, whilst $A_-=\Ct^*k^2\Ct$ is the
extension with Neumann boundary conditions at the origin. The
operators $A_\pm+1$ both have bounded inverse.

The $S$-wave GPI Hamiltonian is defined by
\begin{equation}
\hg = \hat{\Tt}^\dagger\left(
\begin{array}{cc} k^2 & 0 \\ 0 & \Lambda \end{array}\right)
\hat{\Tt}, \label{eq:GPIham}
\end{equation}
where $\Lambda$ is a $\sgn (M_2)$-self-adjoint operator
$\Lambda^\dagger=\Lambda$ on $\MM_2$. To fix $\Lambda$, we require
that $\hg(\psi,0)^T=(-\psi^{\prime\prime},0)^T$ for
all $\psi\in C_0^\infty(0,\infty)$ as a locality requirement.
For general $\psi\in\MM_2$, we have
\begin{equation}
A^*\psi = -\sum_{\omega\in\Omega}\alpha_\omega \omega^2
\ket{\xi_\omega}\inner{\xi_{\overline{\omega}}}{M_2^{-1}\psi},
\end{equation}
so $\MM_2$ is invariant under $A^*$ and it follows immediately from
Section~2.2 that
\begin{Thm} \label{Thm:local}
In the generic case, the unique choice of $\Lambda$ consistent with
locality is
\begin{equation}
\Lambda= -\left(\sgn(M_2) |M_2|^{1/2}\right)^{-1}
\sum_{\omega\in\Omega} \alpha_\omega \omega^2\ket{\xi_\omega}
\bra{\xi_{\overline{\omega}}} |M_2|^{-1/2}.
\end{equation}
\end{Thm}

We proceed to determine the eigenvectors and eigenvalues of
$\Lambda$. First note that
$\inner{\xi_{\overline{\omega_j}}}{M_2^{-1}\xi_{\omega_i}}=
\alpha_{\omega_{i}}^{-1}\delta_{ij}$, which follows from the identity
$\xi_{\omega_i}=\sum\alpha_\omega\ket{\xi_\omega}
\inner{\xi_{\overline{\omega}}}{M_2^{-1}\xi_{\omega_i}}$. It is then
a matter of computation to see that $\varphi_i=\left(\sgn(M_2)
|M_2|^{1/2}\right)^{-1}\xi_{\omega_i}$ is an eigenvector of $\Lambda$
with eigenvalue $-\omega_i^2$ for each $i=1,\ldots,|\Omega|$. Because
$\Lambda$ has rank $|\Omega|$, this exhausts the discrete spectrum of
$\hg$. The following is then immediate.
\begin{Thm}
In the generic case, and  with
$\Lambda$ is defined as above, $\hg$ has the following spectral
properties: $\sigma(\hg)=\sigma_{\rm ac}(\hg) \cup\sigma_{\rm
pp}(\hg)$ where $\sigma_{\rm ac}(\hg)=\RR^+$ and $\sigma_{\rm
pp}(\hg)$ consists of the $|\Omega|$ eigenvalues $-\omega_i^2$, whose
corresponding eigenvectors are
\begin{equation}
\psi_i = \hat{\Tt}^\dagger\varphi_i=
\left(\begin{array}{c} \xi_{\omega_i} \\
\Tt \left(\sgn(M_2) |M_2|^{1/2}\right)^{-1}\xi_{\omega_i}
\end{array}\right).
\end{equation}
The absolutely continuous subspace is the Hilbert space
$\hat{\Tt}^\dagger\HH_k$.
\end{Thm}

This bears out our earlier statement that the poles of the
scattering amplitude on the physical sheet correspond to the discrete
energy spectrum, if locality is imposed.

The physical Hilbert space is required to be a positive definite
invariant subspace of $\Pi_r$ relative to $\hg$.\footnote{An
invariant subspace $\LL$ of a $J$-space $\KK$ relative to a linear
operator $A$ on $\KK$ is a subspace of $\KK$ such that
$\overline{D(A)\cap\LL}=\LL$ and $\Ran A|_{\LL}\subset \LL$, where
the closure is taken in the norm topology of $\KK$.} In $\Pi_k$, we
have the $[\cdot,\cdot]_{\Pi_k}$-orthogonal decomposition
$\Pi_k=\HH_k [+] \MM_2$, where $\MM_2$ is spanned by the eigenvectors
$\varphi_i$ of $\Lambda$. We compute
\begin{equation}
[\varphi_i,\varphi_j]_{\MM_2}
=\inner{\xi_{\omega_i}}{M_2^{-1}\xi_{\omega_j}} \nonumber \\
= \left\{\begin{array}{cl} 0 & \omega_i\not=\overline{\omega_j} \\
\alpha_{\omega_j}^{-1} & \omega_i=\overline{\omega_j}.
\end{array}\right.
\end{equation}
Hence $\Pi_k$ is decomposable as
$\Pi_k = \HH_k [+] E_+ [+] E_- [+] H$
where $E_+$ is spanned by the $\varphi_i$ with
$[\varphi_i,\varphi_i]_{\MM_2}>0$ ($\alpha_{\omega_i}>0$),
$E_-$ is spanned
by those with $[\varphi_i,\varphi_i]_{\MM_2}<0$
($\alpha_{\omega_i}<0$), and
$H$ is the {\em hyperbolic invariant subspace} spanned by those
$\varphi_i$ with $\omega_i\not\in\RR$. Moreover, this is a
decomposition into invariant subspaces, because $D(k^2)$ is dense in
$\HH_k$. The physical Hilbert space $\HH_{\rm phys}$ is therefore
defined by
\begin{equation}
\HH_{\rm phys} = \hat{\Tt}^\dagger (\HH_k[+]E_+).
\end{equation}

We briefly discuss the uniqueness of the GPI Hamiltonian constructed
in this way. As noted in Section~2.1, $\hat{\Tt}$ is unique up to
further unitary dilation and unitary equivalence because the $M_i$
are of finite rank. Further dilation merely corresponds to the
(trivial) freedom to form the direct sum of $\hg$ with the
Hamiltonian of an arbitrary independent system. On the other hand,
replacing $\hat{\Tt}$ by $(\II\oplus U_2)\hat{\Tt}(\II\oplus U_1)$
where $U_i$ is a $\sgn M_i$-unitary operator on $\MM_i$ for $i=1,2$,
it is easy to show that the local GPI Hamiltonian $\hg^\prime$
obtained is given by
\begin{equation}
\hg^\prime=
\left(\begin{array}{cc} \II & 0\\0 & U_1\end{array}\right)^\dagger
\hg\left(\begin{array}{cc} \II & 0\\0 & U_1\end{array}\right).
\end{equation}
We have therefore constructed a family of unitarily equivalent GPI
Hamiltonians on $\Pi_r$ corresponding to the same scattering data. It
is clearly sufficient to study $\hg$ alone in order to determine the
domain and scattering properties of $\hg^\prime$.

\subsection{Domain and Resolvent}

We now determine the domain and explicit action of the operator
$\hg$ under the locality assumption. Our result is the following:
\begin{Thm} \label{Thm:dom}
Let $\Theta_0
=(2/\pi)^{1/2}k^{2\mho-1}(p(k^2)^2+k^2q(k^2)^2)^{-1/2}$. Then in
the generic case,
\begin{eqnarray}
D(\hg)&=&\left\{
\left( \begin{array}{c} \varphi \\ \Phi\end{array}\right)
\mid \varphi,\varphi^\prime\in
AC_{\rm loc}(0,\infty),~\varphi,\varphi^{\prime\prime}\in L^2;\quad
\Phi\in\MM_1, \right.\nonumber \\
&&\qquad\qquad
\left.\begin{array}{c} \ \\ \ \end{array}
\Theta -\lambda[\varphi]\Theta_0 \in
D(k^2)\cap \MM_1\right\} ,
\end{eqnarray}
where $\Theta=\sgn M_1|M_1|^{1/2}\Phi$ and
\begin{equation}
\lambda[\varphi] = \left\{\begin{array}{cl}
P\varphi(0) & \deg p>\deg q \\
P\varphi(0)-Q\varphi^\prime(0) & \deg p=\deg q \\
-Q\varphi^\prime(0) & \deg p< \deg q, \end{array}\right.
\end{equation}
and $P$ and $Q$ are the leading coefficients of $p(z)$ and $q(z)$
respectively. (In the case $M_1=0$,
$D(\hg)=\{\varphi\mid\varphi,\varphi^\prime\in
AC_{\rm loc}(0,\infty),~\varphi,\varphi^{\prime\prime}\in
L^2;~\lambda[\varphi]=0\}$.) Moreover,
\begin{equation}
\hg\left(\begin{array}{c} \varphi \\ \Phi\end{array}\right)=
\left(\begin{array}{c} -\varphi^{\prime\prime}\\ \tilde{\Phi}
\end{array}\right) ,
\end{equation}
where $\tilde{\Phi}$ is given in terms of $\tilde{\Theta}=\sgn M_1
|M_1|^{1/2}\tilde{\Phi}$ by
\begin{equation}
\tilde{\Theta}= k^2(\Theta-\lambda[\varphi]\Theta_0) +
\left(\frac{2}{\pi}\right)^{1/2}
\frac{k(\lambda[\varphi] k^{2\mho} -\varphi(0)p(k^2)
+\varphi^\prime(0)q(k^2))}{(p(k^2)^2+k^2q(k^2)^2)^{1/2}} .
\end{equation}
\end{Thm}
{\em Proof:} The result is a direct application of the discussion in
Section~2.2. The key point is that, for each $\varphi\in
D(-d^2/dr^2|_{C_0^\infty(0,\infty)}^*)$, the vectors $\chi_+$ and
$\chi_-$ are given by
\begin{equation}
\chi_+ = -\varphi(0)e^{-r}\qquad {\rm and} \qquad
\chi_-=\varphi^\prime(0)e^{-r},
\end{equation}
which follows because $\chi_+$ ($\chi_-$) is the unique element of
$\ker (-d^2/dr^2|_{C_0^\infty(0,\infty)}^*+1)$ such that
$\varphi+\chi_+$ ($\varphi+\chi_-$) is in the domain of the Laplacian
with Dirichlet (Neumann) boundary conditions at the origin. $\square$

The resolvent of $\hg$ may be written in the form of Krein's formula
as
\begin{equation}
(\hg-z)^{-1} =
\left(\begin{array}{cc} R_0(z) & 0 \\ 0 & R_1(z)\end{array}\right)
+\frac{q(z)}{p(z)+(-z)^{1/2}q(z)}F(z)F(\overline{z})^\dagger.
\label{eq:reso}
\end{equation}
Here, $R_0(z)=\St^{-1}(k^2-z)^{-1}\St$ is the free resolvent
and the defect element $F(z)\in\Pi_r$ is given by
\begin{equation}
F(z)=\left( \begin{array}{c} e^{-(-z)^{1/2}r} \\
(\sgn M_1|M_1|^{1/2})^{-1}\Psi(z) \end{array}\right),
\end{equation}
where $\Psi(z)\in\MM_1$ is
\begin{equation}
\Psi(z) = \left(\frac{2}{\pi}\right)^{1/2}
\frac{k(p(k^2)q(z)-p(z)q(k^2))}{(k^2-z)(p(k^2)^2+k^2q(k^2)^2)^{1/2}},
\end{equation}
and the operator $R_1(z)$ is defined on $\MM_1$ by
\begin{equation}
R_1(z)\Phi=(\sgn M_1|M_1|^{1/2})^{-1}\left(\frac{2}{\pi}\right)^{1/2}
\frac{k(Q(k^2)-Q(z)q(k^2)/q(z))}{(k^2-z)(p(k^2)^2+k^2q(k^2)^2)^{1/2}},
\end{equation}
where $Q(z)$ is defined in terms of $\Phi$ by
\begin{equation}
\Theta=\sgn M_1|M_1|^{1/2}\Phi=\left(\frac{2}{\pi}\right)^{1/2}
\frac{kQ(k^2)}{(p(k^2)^2+k^2q(k^2)^2)^{1/2}}.
\end{equation}
The above expression for $R(z)$ may be verified directly using
Theorem~\ref{Thm:dom}, and the fact that
\begin{equation}
[(\sgn M_1|M_1|^{1/2})^{-1}\Psi(\overline{z}),\Phi]_{\MM_1}=
-\frac{Q(z)}{q(z)},
\label{eq:clm}
\end{equation}
which is required when one takes inner products with
$F(\overline{z})$. Using this result, it follows that~(\ref{eq:reso})
holds for elements of form $(0,\Phi)^T$ with $Q(z)=0$; direct
computation establishes it for $Q(z)\equiv 1$ and also for  vectors
of form $(\varphi,0)^T$ with $\varphi\in\HH_r$.\footnote{Here, it is
useful to employ the decomposition $\HH_r= \overline{\Ran
(-d^2/dr^2-z)|_{C_0^\infty(0,\infty)}}\oplus \CC
e^{-(-\overline{z})^{1/2}r}$.} Thus~(\ref{eq:reso}) holds on the whole
of $\Pi_r$. It remains to establish equation~(\ref{eq:clm}).
Multiplying through by $q(z)$, the LHS of~(\ref{eq:clm}) is equal to
\begin{equation}
\inner{q(\overline{z})\Psi(\overline{z})}{M_1^{-1}\Theta}=
\inner{\Tt^*q(\overline{z})
\Psi(\overline{z})}{M_2^{-1}\Tt^*\Theta}+\inner{q(\overline{z})
\Psi(\overline{z})}{\Theta}. \end{equation}
Using the identity
$\inner{\xi_{\overline{\omega_j}}}{M_2^{-1}\xi_{\omega_i}}=
\alpha_{\omega_{i}}^{-1}\delta_{ij}$ and the results of Section~3,
the first term is
\begin{equation}
\inner{\Tt^*q(\overline{z})
\Psi(\overline{z})}{M_2^{-1}\Tt^*\Theta}=
\sum_{\omega\in\Omega}\frac{p(z)-\omega
q(z)}{q(-\omega^2)(\omega^2+z)}Q(-\omega^2)\alpha_\omega.
\end{equation}
The required result then follows from the calculation
\begin{eqnarray}
\inner{q(\overline{z})
\Psi(\overline{z})}{\Theta} &=&
\frac{1}{\pi}\int_{-\infty}^\infty dk\frac{ikQ(k^2)
(p(z)-ikq(z))}{(k^2-z)(p(k^2)-ikq(k^2))} \nonumber \\
&=& -Q(z)-\sum_{\omega\in\Omega}\frac{p(z)-\omega
q(z)}{q(-\omega^2)(\omega^2+z)}Q(-\omega^2)\alpha_\omega.
\end{eqnarray}

\subsection{Scattering Theory}
\label{sect:GPIsc}

In this section, we construct M{\o}ller wave operators for $\hg$
relative to the free Hamiltonian $h_0=\St^{-1} k^2\St$ on $\HH_r$ in
order to check that $\hg$ actually exhibits the required scattering
behaviour. Because scattering is a function of the continuous
spectrum only, our results in this section are actually independent
of the precise form of $\Lambda$, and therefore of the locality
requirement.

We work in the $S$-wave, and employ a two space setting: let $B$
be self-adjoint on $\HH_1$, $A$ be self-adjoint on $\HH_2$ and $\JJ$
be a bounded operator from $\HH_1$ to $\HH_2$. Then the M{\o}ller
operators $\Omega^\pm(A,B;\JJ)$ are defined by
\begin{equation}
\Omega^\pm(A,B;\JJ) = \lim_{t\rightarrow\mp\infty}
e^{iAt}\JJ e^{-iBt}P_{\rm ac}(B),
\end{equation}
and are said to be complete if the closure of ${\rm
Ran}\Omega^\pm(A,B;\JJ)$ is equal to ${\rm Ran} P_{\rm ac}(A)$.

In the following,
$\JJ_r$ and $\JJ_k$ are the natural embeddings of $\HH_r$ and
$\HH_k$ into $\Pi_r$ and $\Pi_k$ respectively.

\begin{Thm} Let $\JJ:\HH_r\rightarrow\Pi_r$ be given by
$\JJ=\hat{\Tt}^\dagger \JJ_k\Tt$. Then
$\Omega^\pm(\hg,h_0;\JJ)$ exist, are complete, and given by
\begin{equation}
\Omega^\pm(\hg,h_0;\JJ) = \hat{\Tt}^\dagger \JJ_k
e^{\pm i\delta_0(k)}\St,
\label{eq:Mops}
\end{equation}
where $\delta_0(k)$ is given by~(\ref{eq:GPIlow}).
 \end{Thm}
{\em Proof:} Writing $U_t$ for multiplication by $e^{-ik^2t}$ on
$\HH_k$, we have
\begin{eqnarray}
e^{i\hg t} \JJ e^{-ih_0t}P_{\rm ac}(h_0) &=& \hat{\Tt}^\dagger
\left(\begin{array}{cc} U_{-t} & 0 \\ 0 & \exp i\Lambda t
\end{array}\right) \hat{\Tt} \JJ  \St^{-1} U_t \St
\nonumber \\
&=& \hat{\Tt}^\dagger
\JJ_k U_{-t} \Tt  \St^{-1} U_t \St.
\end{eqnarray}
Now, for any $u(k)\in C_0^\infty(0,\infty)$,
\begin{eqnarray}
\| U_{-t}\Tt\St^{-1}U_{t}u(k)-e^{\pm i\delta_0(k)} u(k)\|^2 & = &
\|\sin\delta_0(k)\Ct (\Ct^{-1}\pm i\St^{-1})U_t u(k)\|^2
\nonumber\\ & \le & \frac{2}{\pi}\int_0^\infty dr
\left|\int_0^\infty dk e^{i(\pm kr-k^2t)} u(k)\right|^2,
\end{eqnarray}
which vanishes as $t\rightarrow\mp\infty$ by (non)-stationary phase
arguments (see the Corollary to Theorem XI.14 in \cite{RSiii}). Thus
$U_{-t}\Tt\St^{-1}U_t \rightarrow e^{\pm i\delta_0(k)}$ strongly as
$t\rightarrow\mp\infty$. The existence and form of the M{\o}ller
operators are then immediate. One easily checks that they are unitary
maps from $\HH_r$ to $P_{\rm ac}(\hg)=\hat{\Tt}^\dagger\JJ_k\HH_k$,
to establish completeness. $\square$

We conclude that our construction does indeed yield the required
scattering theory, and also that -- as a by-product of the
construction -- complete M{\o}ller operators may easily and
explicitly be determined.

\sect{Examples}

As an application, we construct the class of GPI models with
scattering data
\begin{equation}
\cot\delta_0(k) = -\frac{1}{kL}+kM, \label{eq:ERlow}
\end{equation}
where $L$ is the scattering length, and $M$ is twice the effective
range. These models therefore represent the effective range
approximation to the behaviour of a non-point interaction in the
$S$-wave. This class of models has been partially studied by
Shondin~\cite{Shond1}, who considered the case $M<0$ (`models of type
$B_2$') and also appears as a special case of the models considered
by Pavlov in \cite{Pav1}.  (We also note that van Diejen and Tip
\cite{Diejen} have constructed models of type $\cot\delta_0(k) = (ak
+ bk^3+ck^5)^{-1}$ using the distributional method.) The case $M>0$
does not appear to have been treated before. Our construction
provides a unified construction for all models in the above class,
and also provides the spectral representation such models as a
by-product of the construction (although we will not state this
explicitly).

The above class of GPI models contains two interesting
sub-families: the ordinary point interactions ($M=0$) and also the
resonance point interactions arising formally by setting $L=\infty$,
i.e., $\cot\delta_0(k)=kM$ with $M\in\RR\cup\{\infty\}$. Such models
are required in situations where the scattering length is
generically forced to be infinite, for example in certain systems of
supersymmetric quantum mechanics.

We begin by briefly treating the point interactions, both for
completeness and also to demonstrate how this class arises in our
formalism. We then turn to the general case, obtaining RPI models
in the limit $L\rightarrow -\infty$.

\subsection{Point Interactions}

The required integral transform is
\begin{equation}
\Tt = (1+(kL)^2)^{-1/2}\St - kL(1+(kL)^2)^{-1/2}\Ct.
\end{equation}
In the cases $L=0,\infty$, $\Tt$ reduces to $\St$ and $\Ct$
respectively, and the Hamiltonian is given immediately by
$\Tt^*k^2\Tt$. We exclude these cases from the rest of our discussion.

We therefore apply the construction of Section 3, with $p(z)\equiv
-L^{-1}$ and $q(z)\equiv 1$. We find that $\mho=0$, so $M_1=0$ (i.e.,
$\Tt\Tt^*=\II$). Straightforward application of
Proposition~\ref{Prop:M2} yields
\begin{equation}
M_2=\left\{\begin{array}{cl}
\ket{\chi_L}\bra{\chi_L} & L>0 \\ 0 & L<0, \end{array}\right.
\end{equation}
where $\chi_L(r)=(2/L)^{1/2}e^{-r/L}$ is normalised to unity.
Hence if $L<0$, $\Tt$ is unitary and the Hamiltonian is
$h_L = \Tt^* k^2 \Tt$,
with purely absolutely continuous spectrum $\RR^+$. In the case
$L>0$, the momentum Hilbert space is extended to $\HH_k\oplus\CC$,
representing a single bound state, and the unitary dilation
$\hat{\Tt}:\HH_r\rightarrow\HH_k\oplus\CC$ takes form
\begin{equation}
\hat{\Tt} =
\left(\begin{array}{c} \Tt \\ \bra{\chi_L} \end{array}\right);
\qquad \hat{\Tt}^*=
\left(\begin{array}{cc} T^* & \ket{\chi_L}\end{array}\right).
\end{equation}
($\HH_k\oplus\CC$ has the obvious inner product.)
The Hamiltonian is
\begin{equation}
h_L = \hat{\Tt}^{-1}\left(\begin{array}{cc} k^2 & 0 \\ 0 & \lambda
\end{array}\right) \hat{\Tt} = \Tt^* k^2\Tt +
\lambda\ket{\chi_L}\bra{\chi_L},
\end{equation}
and the locality requirement fixes $\lambda= -L^{-2}$, which is, of
course, the usual value. Finally, the domain of $h_L$ is given by
Theorem~\ref{Thm:dom} as the space of $\varphi$ with
$\varphi,\varphi^\prime\in AC_{\rm loc}(0,\infty)$,
$\varphi^{\prime\prime}\in L^2$ and satisfying
the well known boundary condition
\begin{equation}
\varphi(0)+L\varphi^\prime(0) = 0.
\end{equation}
{}To summarise, all the well known properties of point interactions
may be derived within our formalism.

\subsection{Effective Range Approximation}

In this section, we maintain $M\not=0$, $L\not=0$, setting
$p(z)=-L^{-1}+zM$ and $q(z)\equiv 1$. We will not explicitly
construct the dilation (although this follows immediately from our
discussion), but will use the results of Section~4 to read off the
domain and action of the GPI Hamiltonian $h_{L,M}$.

Using the results of Section~3, we find
\begin{equation}
\mho=1;\qquad |\Omega|=
\left\{
\begin{array}{cl} 1+\frac{1}{2}(\sgn M+\sgn L) & L\not=\infty \\
\frac{1}{2}(1+\sgn M) & L=\infty. \end{array}
\right.
\end{equation}
Writing $W(z)=-M(z-\omega_1)(z-\omega_2)$, $\Omega$ is
the subset of $\{\omega_1,\omega_2\}$ lying in the left-hand
half-plane, and we have $\omega_1+\omega_2=-M^{-1}$,
$\omega_1\omega_2=(ML)^{-1}$. The residues $\alpha_\omega$ are
\begin{equation}
\alpha_{\omega_1}=-\frac{2\omega_1}{M(\omega_1-\omega_2)},\qquad
\alpha_{\omega_2}=\frac{2\omega_2}{M(\omega_1-\omega_2)}.
\end{equation}
In addition, the space $\MM_1=\Ran M_1$ is equal to $\CC\ket{\eta}$,
where
\begin{equation}
\eta(k) = {\cal N} \frac{k}{(k^2+(k^2M-L^{-1})^2)^{1/2}},
\end{equation}
and the normalisation constant is
\begin{equation}
{\cal N}=\left\{\begin{array}{cl}
(2|M|/\pi)^{1/2} & ML>0 \\
(2|M|/\pi)^{1/2}(1-4ML^{-1})^{1/4} & ML<0. \end{array}
\right.
\end{equation}
Using Proposition~\ref{Prop:M1}, we obtain
\begin{equation}
M_1 = \lambda\ket{\eta}\bra{\eta} ;
\qquad \lambda = \left\{\begin{array}{cl}
+1 & M<0, L<0 \\
-\sgn M (1-4ML^{-1})^{-1/2} & ML<0 \\
-1 & M>0, L>0.
\end{array}\right.
\end{equation}

Accordingly, the extended position inner product space is
$\Pi_r=\HH_r\oplus\CC$ with $J$-inner product specified by
$J=\II\oplus (-\sgn M)$. The scalar component is the coefficient of
$\ket{\eta}$ in $\MM_1$. For all generic cases (i.e., all cases
other than $L=4M>0$) Theorem~\ref{Thm:dom} entails that the domain
of $h_{L,M}$ is
\begin{equation}
D(h_{L,M}) = \left\{\left(\begin{array}{c} \varphi \\ \Phi
\end{array}\right) \mid \varphi,\varphi^\prime\in AC_{\rm
loc}(0,\infty),~\varphi,\varphi^{\prime\prime}\in L^2;\quad \Phi =
-|M|^{1/2}\varphi(0)  \right\},
\label{eq:DhLM}
\end{equation}
and that the action is
\begin{equation}
h_{L,M} \left(\begin{array}{c} \varphi \\ -|M|^{1/2}\varphi(0)
\end{array}\right) =
\left(\begin{array}{c}
-\varphi^{\prime\prime}
\\ -\sgn M |M|^{-1/2}(\varphi^\prime(0)+L^{-1}\varphi(0))
\end{array}\right).
\end{equation}
Moreover, one may show that these equations also hold in the
non-generic case $L=4M>0$.

It is worth noting how this domain and action correspond to the
scattering data~(\ref{eq:ERlow}). Solving the equation
$h_{L,M}(\varphi,\Phi)^T=k^2(\varphi,\Phi)^T$ for the generalised
eigenfunctions of $h_{L,M}$, we find
$\varphi(r)\propto\sin(kr+d(k))$ for some $d(k)$, and also obtain the
relation
\begin{equation}
-\sgn M |M|^{-1/2}(\varphi^\prime(0)+L^{-1}\varphi(0))=
-k^2|M|^{1/2}\varphi(0),
\end{equation}
which entails that $k\cot d(k) = \varphi^\prime(0)/\varphi(0) =
-L^{-1}+k^2M$. Thus $d(k)$ is precisely the scattering data
$\delta_0(k)$.

The RPI models, which have scattering data $\cot\delta_0(k) =kM$ are
obtained in the same way. The space $\MM_1$ is spanned by
$\psi_M(k) =(2|M|/\pi)^{1/2}(1+(kM)^2)^{-1/2}$, and the operator
$M_1$ is found to be $M_1=-(\sgn M)\ket{\psi_M}\bra{\psi_M}$. Thus
the inner product space is $\Pi_r=\HH_r\oplus\CC$ with
$J=\II\oplus(-\sgn M)$. They have the domain~(\ref{eq:DhLM}) and
action
\begin{equation}
h_{{\rm rpi},M} \left(\begin{array}{c} \varphi \\ -|M|^{1/2}\varphi(0)
\end{array}\right) =
\left(\begin{array}{c}
-\varphi^{\prime\prime}
\\ -\sgn M |M|^{-1/2}\varphi^\prime(0)
\end{array}\right).
\end{equation}

Let us consider the physical Hilbert space for these models. From
Section~4, this is constructed by projecting out the hyperbolic
invariant subspace, and also those eigenfunctions with negative norm
squared (if present). The bound states of $h_{L,M}$ are clearly
vectors of form $(\xi_\omega,|M|^{1/2})^T$ with norm squared equal to
$-(2{\rm Re}\,\omega)^{-1}-M$, where $\omega$ is a root of
$\omega^2+M^{-1}\omega+(ML)^{-1}=0$. There are four cases to consider:

{\noindent \em Case (i): $M<0$}. $\Pi_r$ is positive definite so no
projection is required.

{\noindent \em Case (ii): $M>0$, $L<0$}. There is a unique bound
state with
\begin{equation}
\omega=\frac{1+(1-4M/L)^{1/2}}{-2M}
\label{eq:ome}
\end{equation}
and negative norm squared. Projecting this state out, we obtain
\begin{equation}
\HH_{\rm phys} =
\left\{\left(\begin{array}{c} \varphi \\
M^{-1/2}\inner{\xi_\omega}{\varphi}\end{array}\right)\mid
\varphi\in\HH_r \right\}.
\label{eq:HP}
\end{equation}

{\noindent \em Case (iii): $M>0$, $0<L<4M$}. There are two bound
states with complex conjugate eigenvalues. Accordingly, their
eigenfunctions span a hyperbolic invariant subspace. Projecting
this subspace out, we find
\begin{equation}
\HH_{\rm phys} =
\left\{\left(\begin{array}{c} \varphi \\
M^{-1/2}\inner{
\frac{1}{2}(\xi_\omega+\xi_{\overline{\omega}})}{\varphi}
\end{array}\right)\mid\varphi\in\HH_r~{\rm s.t.}~
\inner{(\xi_\omega-\xi_{\overline{\omega}})}{\varphi}=0 \right\},
\end{equation}
where $\omega$ is given by equation~(\ref{eq:ome}).

{\noindent \em Case (iv): $M>0$, $L>4M$}. There are two bound
states with real eigenvalues. However, only the state specified
by~(\ref{eq:ome}) has negative norm. Projecting this out, we arrive
at the same expression for $\HH_{\rm phys}$ as in case (ii).

RPI models are covered by Case (i) for $M<0$, and have $\HH_{\rm
phys}$ given by~(\ref{eq:HP}) for $M>0$, with $\omega=-1/M$.

The GPI Hamiltonian acts on $\HH_{\rm phys}$ by restriction. For
example, in case (ii) above, we have
\begin{eqnarray}
D(h_{L,M}|_{\HH_{\rm phys}}) &=& \left\{
\left(\begin{array}{c} \varphi \\
M^{-1/2}\inner{\xi_\omega}{\varphi}\end{array}\right)\mid
\varphi,\varphi^\prime\in AC_{\rm loc}(0,\infty),~
\varphi,\varphi^{\prime\prime}\in L^2;\right. \nonumber \\
&&\qquad\qquad\qquad\qquad\quad
\left.\begin{array}{c} \ \\ \ \end{array}
M\varphi(0)=-\inner{\xi_\omega}{\varphi} \right\}
\end{eqnarray}
on which $h_{L,M}|_{\HH_{\rm phys}}$ acts as before. The restricted
operator has the same continuum spectrum as $h_{L,M}$, but has no
bound states in this case. Moreover, the property of locality is
partially lost: it is clear that vectors of form $(\varphi,0)^T$
with $\varphi\in C_0^\infty(0,\infty)$ are in $\HH_{\rm phys}$ only if
$\varphi\perp\xi_\omega$. However, for elements of this form in
$\HH_{\rm phys}$, it remains the case that $h_{L,M}|_{\HH_{\rm
phys}}(\varphi,0)^T= (-\varphi^{\prime\prime},0)^T$. Thus the
properties of locality and `positivity' are not entirely compatible.

\subsection{Physical Interpretation}

In this section, we discuss how the effective range models
constructed above may be used to model Schr\"{o}dinger operators
$H=-\triangle+V$, where $V$ is smooth, spherically symmetric and
compactly supported within radius $a$ of the origin. Our methodology
extends that described in~\cite{KF}, in which the scattering length
approximation is discussed.

Given a smooth spherically symmetric potential $V(r)$ supported
within radius $a$ of the origin, we may find the `best fit' GPI model
$h_{L,M}$ as follows. Let $u_0$ be the $S$-wave zero energy
eigenfunction, i.e., the solution to $-u_0^{\prime\prime}+Vu_0=0$
with regular boundary conditions at the origin.  Then the arguments
of Section 11.2 of \cite{Newt} give the low energy parameters $L$ and
$M$ as
\begin{equation}
L= a - \left.\frac{u_0}{u_0^\prime}\right|_{r=a};
\label{eq:L}
\end{equation}
and
\begin{equation}
M = a\left\{ 1-\frac{a}{L}
+\frac{1}{3}\left(\frac{a}{L}\right)^{2}-
\left( 1-\frac{a}{L}\right)^2
\frac{\int_0^a |u_0(r)|^2 dr}{a|u_0(a)|^2} \right\}.
\label{eq:M}
\end{equation}
Thus the scattering behaviour is $\cot\delta_0(k)=
-(kL)^{-1}+kM+O(k^3)$ and the best fit GPI model in our class is
$h_{L,M}$. We refer to equations~(\ref{eq:L}) and~(\ref{eq:M}) as
{\em fitting formulae}; equation~(\ref{eq:L}) is the fitting formula
employed in \cite{KF}. The range of energies for which the
approximation is valid can be determined by a `believability'
analysis analogous to that described in \cite{KF}. We will not do
this here.

Note that $M$ obeys the bound
\begin{equation}
-\infty \le M < a\left\{ 1-\frac{a}{L}
+\frac{1}{3}\left(\frac{a}{L}\right)^{2}\right\}.
\end{equation}
Moreover, this bound is best possible: for any
$L\in\RR\cup\{\infty\}$ and any $M$ in the above range, one can
clearly find a smooth function $u_0(r)$ satisfying regular boundary
conditions at the origin, $u_0\propto(1-r/L)$ for $r>a$ and such
that~(\ref{eq:M}) holds. Then the potential defined by
$V(r)=u_0^{\prime\prime}(r)/u_0(r)$ has $S$-wave scattering behaviour
approximated to second order by $h_{L,M}$. The contribution to the
total scattering cross section from the effective range term
generally outweighs that from higher angular momenta, so the $S$-wave
GPI model provides a second order approximation to the full
scattering behaviour.

Finally, we discuss the interpretation of the discrete spectrum of
$h_{L,M}$. We have constructed $h_{L,M}$ so that its
scattering behaviour matches that of a given Schr\"{o}dinger
operator at low energies, $E$. For larger $|E|$, the approximation
breaks down -- in the language of \cite{KF} we say that it is no
longer `believable'. Thus, deeply bound states are unlikely to be
believable. In particular, for $0<L<4M$, $h_{L,M}$ exhibits a complex
conjugate pair of eigenvalues, which can never be
believable.\footnote{These are not eigenvalues of $h_{L,M}$
restricted to the physical Hilbert space. However, they persist as
poles in the scattering amplitude and our remarks still apply:
$h_{L,M}$ does not give a reliable approximation to the scattering
theory at those scales.} Such phenomena are artifacts of the
idealisation process, due to the truncation of the low energy
expansion. The issue of believability is discussed in \cite{KF};
similar comments are made in \cite{Diejen}.

\sect{Conclusion}

We begin by discussing various generalisations of our method. There
are many situations in which the analysis of Section~2.2 may be
applied. In two dimensions, for example, one can consider radial GPI
Hamiltonians which agree with
\begin{equation}
h = -\frac{1}{r}\frac{d}{dr}r\frac{d}{dr} + \frac{\nu^2}{r^2}
\end{equation}
away from the origin, as models for an infinitesimal `dot' of
magnetic flux $\nu$, with $|\nu|<1$. In this case, one must employ
Hankel transforms rather than sine and cosine transforms. In \cite{F}
we will implement this programme to construct a models of RPI type
for the Dirac equation in the presence of an infinitesimal tube of
flux. These models provide the leading order approximation to the
scattering data.

Our method could also be applied to $S$-wave GPI models with a
Coulombic tail. In this case, the appropriate integral transforms
would be based upon Whittaker functions and the scattering data
would be specified in terms of Coulomb-modified partial wave shifts.
In this case, the dimension of $\MM_2$ would be countably
infinite, due to the countable discrete spectrum of
such models. However, one would expect $\MM_1$ to remain finite
dimensional for simple models.

Secondly, it is of interest to generalise the unitary dilation method
to sectors of higher angular momentum with $\ell\ge 1$ (and the
corresponding analogues for magnetic flux dots -- i.e., $|\nu|\ge 1$
-- and Coulombic GPI for $\ell\ge 1$). This is more problematic,
because the radial Hamiltonian $-d^2/dr^2+\ell(\ell+1)/r^2$ is
essentially self-adjoint on $C_0^\infty(0,\infty)$ and so the method
of Section~2.2 does not apply. Here, it might be possible to  obtain
a suitable integral transform by analysing the distributional
construction. We hope to return to this elsewhere.

Finally, we consider applications to the definition of arrays of
point scatterers. Here, the most likely use of our methods is to
generate the `monomer' by inverse scattering. By passing to the
resolvent written in the form of Krein's formula, one can isolate
the appropriate `defect element' and proceed to form the array by
methods discussed in \cite{Diejen}, which generalise the procedure
for arrays of PI developed in \cite{Gmann}.

{}To summarise, we have introduced an inverse scattering construction
for GPI models using the theory of unitary dilations, and developed
the method in detail for the class of single centre $S$-wave GPI
models with rational $S$-matrices. A physical locality requirement
completes the specification of the Hamiltonian, whose scattering,
spectral and domain properties are explicitly determined from our
results.

{\em Acknowledgements:} The notions of fitting formulae and the
general methodology of Section~5.3 are due in origin to Bernard Kay
\cite{KF}; I thank him and also Graham Allan and Clive Wells for
useful conversations. In addition, I thank Churchill College,
Cambridge, for financial support under a Gateway Studentship.


\end{document}